\documentclass[prl,twocolumn,superscriptaddress,showpacs,floatfix]{revtex4-2}
\usepackage{graphicx}
\usepackage{dcolumn}
\usepackage{braket}
\usepackage{bm}
\usepackage{amsfonts} 
\usepackage{amsmath} 
\usepackage[bookmarksnumbered,pdfpagelabels=true,plainpages=false,colorlinks=true,linkcolor=blue,citecolor=blue,urlcolor=blue]{hyperref}
\usepackage[bookmarksnumbered,pdfpagelabels=true,plainpages=false,colorlinks=true,linkcolor=blue,citecolor=blue,urlcolor=blue]{hyperref}

\def \usach {Departamento de F\'isica, Universidad de Santiago de Chile, 9170124, Santiago, Chile.}
\def \cedenna {Centro  de Nanociencia y Nanotecnología CEDENNA, Avda. Ecuador 3493, Santiago, Chile.}
\def \fcfm {Departamento de F{\'i}sica, Facultad de Ciencias Físicas y Matemáticas, Universidad de Chile, Santiago, Chile.}

\def \utarapaca {Departamento de Física, Facultad de Ciencias, Universidad de Tarapacá, Casilla 7-D, Arica, Chile.}

\begin{document}

 \title{Electric-Field Control of Josephson Oscillations in Dipolar Bose-Einstein Condensates}

\author{David Galvez-Poblete}
\email{david.galvez.p@usach.cl}
\affiliation{\usach}
\affiliation{\cedenna}

\author{Roberto E. Troncoso}
\affiliation{\utarapaca}

\author{Guillermo Romero}
\affiliation{\usach}
\affiliation{\cedenna}

\author{Alvaro S. Nunez}
\affiliation{\fcfm}

\author{Sebastian Allende}
\affiliation{\usach}
\affiliation{\cedenna}

\begin{abstract}
We study the dynamic behavior of a Bose-Einstein condensate (BEC) with dipolar interactions when the influence of external electric fields affects the coherent tunneling properties. Here, we propose a tunable platform based on BECs where Josephson oscillations can be engineered and modulated through external electric fields. We develop a theoretical and numerical framework that reveals how electric fields affect intercondensate tunneling, phase dynamics, and collective excitations. By employing a coupled set of Gross–Pitaevskii equations with adiabatic elimination of excited states, we demonstrate field-induced tuning of Josephson frequencies and a transition from contact to dipole-dominated regimes. These findings corroborate theoretical predictions about the sensitivity of dipolar BECs to external fields and deepen our understanding of quantum coherence and tunneling in long-range interacting quantum systems.
\end{abstract}

\maketitle


\section{INTRODUCTION}

Exploring quantum mechanical effects in macroscopic systems offers profound insights into the fundamental aspects of quantum coherence and tunneling. Bose-Einstein condensates (BECs) are an ideal platform for investigating these phenomena because of their unique quantum mechanical nature \cite{Annett2004,Pethick2008}. One of the clearest manifestations of quantum coherence \cite{Yang1962} in such systems is the Josephson effect, where a supercurrent flows across two weakly coupled condensates \cite{Levy2007, PhysRevA.59.620, Giovanazzi2000,Sakmann2011,Leggett2006,Valtolina2015,Adhikari2014,Abad2011}. Predicted by Brian Josephson in 1962 in the context of superconductivity, this quantum mechanical phenomenon arises from the coherent tunneling of Cooper pairs, offering a mesmerizing manifestation of the broken symmetry state associated with superfluidity \cite{Josephson1962}.  The effect is fundamental in superconducting quantum circuits, SQUIDs, and quantum computing technologies \cite{Makhlin2001, Paik2011}. A Bose-Einstein condensate (BEC) of dipolar molecules offers several remarkable advantages over traditional BECs of atoms with short-range interactions: (a) Long-range and anisotropic interactions: Dipolar interactions extend beyond their nearest neighbors and depend on the relative orientation of the dipoles, allowing novel quantum phases and tunable interaction geometries not possible in contact-interaction BECs \cite{Yi2000,Tikhonenkov2008,Gral2000,Blakie2020,Diniz2020,Lahaye_2009, Fischer2006, Ronen2006, Abad2010}. (b) Rich many-body physics: These systems can host exotic phases like supersolids \cite{Norcia2021,SnchezBaena2024,Kirkby2024,Trypogeorgos2025}, quantum ferrofluids \cite{Lahaye2007,Wilson2012}, and roton-like excitations \cite{Ronen2007, Wilson2008,Bisset2013}, enriching the landscape of quantum simulation. (c) Quantum simulation of complex models: The controllable nature of dipolar interactions allows simulation of extended Hubbard models \cite{Greiner2002,Baier2016}, spin-lattice systems \cite{Yan2013}, and lattice gauge theories with potential applications in quantum magnetism and topological order \cite{Lin2011,Goldman2014,Yao2022}. (d) Enhanced tunability: External electric or magnetic fields can precisely control the strength and orientation of dipolar interactions, offering fine-grained control over quantum dynamics. (e) Access to strong correlation regimes: The interplay between long-range order and quantum fluctuations in dipolar BECs makes them a powerful platform for exploring quantum phase transitions \cite{Xu2024}, many-body localization \cite{Zhang2015}, and non-equilibrium phenomena.
In short, BECs of dipolar molecules open a pathway to study strongly correlated, long-range interacting quantum systems in a highly controllable setting.

Recent experimental breakthroughs, as detailed in Ref.~\cite{Bigagli2024}, have successfully realized BECs from dipolar molecules. This challenging feat enables the direct observation of their intrinsic properties under controlled conditions. This achievement provides a robust platform to test theoretical predictions and to probe dipolar interactions in regimes beyond the contact-interaction limit. These interactions alter the traditional Josephson dynamics observed in non-dipolar BECs, leading to new phenomena such as modified tunneling rates and phase stability, which are crucial for applications in quantum simulation and information processing \cite{Byrnes2012Apr,Byrnes2015Feb}. Furthermore, manipulating these interactions through external fields has been shown to tune the characteristics of the Josephson junctions, such as their oscillation frequency and amplitude, offering a method to control macroscopic quantum states dynamically \cite{Lahaye_2009}. Related approaches have also demonstrated that measurement backaction, such as Faraday-imaging, can induce squeezing in double-well BECs \cite{IloOkeke2021}.

In this work, we build upon the foundational studies discussed in \cite{Bigagli2024} to explore the dynamical effects in dipolar BECs, focusing on how external electric fields can be used to modulate the Josephson oscillation frequencies. By aligning the dipole moments of the particles within our BECs and varying the electric field parameters, we systematically study their impact on the Josephson junction properties. Our results reveal that the oscillation frequencies can be finely tuned by adjusting the field strength and orientation, reflecting changes in dipole-dipole interaction energies and inter-condensate phase differences. These findings corroborate theoretical models that predict the sensitivity of dipolar BECs to external fields and enhance our understanding of quantum coherence and tunneling in complex quantum systems. Our results have significant implications for developing quantum sensors and simulators that leverage the tunable nature of dipolar interactions in BECs.


\textit{A phenomenological theory of dipolar BECs}: To approach this work, it is first necessary to study the nature of a Bose-Einstein condensate (BEC) that exhibits dipolar interactions. As a starting point, we consider the work conducted by Bigagli \textit{et al.}~\cite{Bigagli2024}, in which the stabilization of a BEC of NaCs molecules was achieved by coherently coupling three states: $\ket{J, m_J} = \ket{0,0}$ (a state with no dipole moment), $\ket{1,0}$ (a state with a dipole moment oscillating out of the $xy$-plane), and $\ket{1,1}$ (a state with a rotating dipole moment within the $xy$-plane). The stability of the condensate is attributed to the combination of two microwave fields: one with circular polarization ($\sigma^+$), which induces in-plane rotating dipole moments and short-range repulsive interactions; and another with linear polarization ($\pi$), which induces vertically oscillating dipole moments that lead to long-range attractive interactions. The controlled superposition of these effects allows for the compensation of long-range attraction while preserving the short-range repulsion, resulting in a net repulsive effective potential, which is essential for the stability of the BEC. Based on this experimental result, we model a dipolar BEC using classical fields as order parameters corresponding to each of the rotational states ($\Psi_1 \to\ket{0,0}$, $\Psi_2 \to \ket{1,0}$, and $\Psi_3 \to \ket{1,1}$). The energy difference between the ground state $|0,0 \rangle$ and the excited states $|1,0 \rangle$ and $|1,1 \rangle$ is on the order of 3.471 GHz. Also, Rabi frequencies and detunings are on the order of 10 MHz.

To study this system, we consider that in the first condensate ($\Psi_1$), only self-interaction between the molecules is present, while in the second ($\Psi_2$) and third ($\Psi_3$) condensates, both dipolar and self-interactions must be taken into account. Additionally, we included the cross-interactions between the condensates, considering that all three condensates share the same physical space. Based on these considerations, we derive the following generalized Gross-Pitaevskii (GP) matrix equation \cite{Pitaevskii2016, Fadel2021, Troncoso2011, Troncoso2014}:
\begin{equation}\label{eq: Gross-Pitaevski}
    i \hbar \partial_t |\Psi\rangle = [\mathbb{H}_o+ \mathbb{B}  + \gamma \langle \Psi |\Psi \rangle ]|\Psi \rangle,
\end{equation}
where $|\Psi \rangle = ( \Psi_1, \Psi_2, \Psi_3)$ is a three-component spinor containing the classical field of each condensate. $\mathbb{H}_o$  represents the Hamiltonian matrix without coupling terms among the condensates and the contact interaction. $\mathbb{B}$ is a matrix that models the external pumping terms. These matrices, whose deduction is detailed at the Supplemental Material (SM), are given by:

\begin{equation}
\mathbb{H}_o = -\dfrac{\hbar^2}{2m} \nabla^2 
+
\begin{pmatrix}
V & 0 & 0 \\
0 & V  + \Phi_2 + \hbar \Gamma & 0 \\
0 & 0 & V  + \Phi_3 + \hbar \Gamma
\end{pmatrix}
\end{equation}

\begin{equation}
    \mathbb{B} = \begin{pmatrix}
        0 & \nu_\pi^* & \nu_\sigma^*\\
        \nu_\pi & 0 & 0 \\
        \nu_\sigma & 0 &0
    \end{pmatrix}. 
\end{equation}

 Here, $m$ represents the molecular mass of NaCs. $\mathbb{I}$ is the matrix unit. $V$ corresponds to the external potential acting on each condensate. In general, this is a harmonic trap common to all three condensates; still, additional terms, such as electric field interactions or external potential barriers, can also be included. $\Phi_j$ represents the dipolar interaction term and is given by \cite{Lima2012,Pedri2005}:
\begin{equation*}
    \Phi_j = \int d\textbf{r}' V_{ddj}(\vec{r}-\vec{r}') |\Psi_j (\vec{r}')|^2  ; \quad V_{ddj} = \frac{C_{ddj}}{4 \pi } \frac{1 - 3\cos^2(\theta_j)}{|\vec{r}-\vec{r}'|^3},
\end{equation*}
with $C_{ddj}$ denoting the dipolar interaction strength, which depends on the electric dipole moment of each condensate. $\theta_j$ is the angle between the relative position vector connecting two dipoles and the orientation of the electric dipole moment. $\hbar \Gamma$ corresponds to the energy difference between the ground state and the two excited states. The significant energy that separates the first condensate with the second and third one, allows us to assume that the ground state is considerably more populated than the $\Psi_2$ and $\Psi_3$ states. Finally, the terms $\nu_\pi = \hbar \Omega_\pi e^{-i t \Delta }$ and $\nu_\sigma = \hbar \Omega_\sigma e^{-i t \Delta }$ represent the pumping that couple $\Psi_1 $ with $\Psi_2$, and $\Psi_1$ with $\Psi_3$, respectively. $\Omega_\pi $ and $\Omega_\sigma$ are the Rabi frequencies of the system, and $\Delta$ is the detuning frequency, which we assume, for simplicity, to be the same for both coupling signals. As a result, the contribution of the upper states to the total density is minimal (without an external electric field), and their influence on the system manifests only through minor corrections. Therefore, Eq. (\ref{eq: Gross-Pitaevski}) is the essential starting point for the discussion presented below.

In the present work, we primarily aim to study the stationary and dynamical behavior of a multicomponent dipolar BECs, as a preliminary step toward analyzing the Josephson effect in the system and its properties under an applied electric field.
The following calculations were performed under the assumption of a two-dimensional (2D) system, considering $m = 2.588 \times10^{-25} kg$, $ \Gamma = 2\pi \times 3.471 $ GHz, $\Omega_\pi = 2\pi  \times 6.5 $ MHz, $\Omega_\sigma = 2\pi \times 7.9$ MHz, and $\Delta = 2 \pi  \times 10 $ MHz \cite{Bigagli2024}. In addition, an isotropic harmonic trap in the $XY$ plane with a frequency of $\omega_{t} = 50$ Hz was considered. We also used the following value for the self-interaction parameter $\gamma = 6.71 \times 10^{-44} J\cdot m^2$. To obtain these values, we used the experimentally observed scattering lengths and calculated the three-dimensional self-interaction parameter as $\gamma_{3D} = {4 \pi \hbar^2 a}/{m}$ \cite{Pitaevskii2016}. Then, we integrated over the tightly confined third dimension to obtain the effective two-dimensional self-interaction parameter. For this calculation, we consider the system is confined in the $z$-direction \cite{Grlitz2001} by a harmonic trap with a frequency of $ 10$ kHz. The same procedure was followed to obtain $C_{dd} =  {12 \pi \hbar^2 a_d}/{m}$, taking into account the reported dipolar scattering length $a_d$ \cite{Lahaye_2009}.

To compute the dipolar interaction terms, $\Phi_2$ and $\Phi_3$, we first obtained the effective two-dimensional Fourier transform of each potential (see SM for details), $\tilde{V}_{2,dip}^{2D}(k) = \frac{C_{dd}}{2\pi}[ \frac{2\sqrt{2\pi}}{3 l_z}-\pi k e^{k^2 l_z^2 /2} {\rm erfc}(k l_z /\sqrt2)]$ and $\tilde{V}_{3,dip}^{2D}(k) = \frac{C_{dd}}{2\pi}[\frac{k \pi}{2} e^{k^2 l_z^2 /2} {\rm erfc}(k l_z /\sqrt2) - \frac{\sqrt{2\pi}}{3 l_z}]$, with $l_z$ denoting the characteristic confinement length along the $z$ direction, ${\rm erfc}$ is the complementary error function and $k = \sqrt{k_x^2+k_y^2}$. Subsequently, we performed the inverse Fourier transform to evaluate these terms in real space.

Before proceeding, the first step is to eliminate time-dependent terms in the pumping by the transformation $\Psi_j = \psi_j e^{it \Delta_j }$, see SM for details. Then, we rewrote Eq.~(\ref{eq: Gross-Pitaevski}) in dimensionless form by rescaling the variables using the trapping potential and normalized the total wave function according to $\int d^2r (|\Psi_1 |^2 + |\Psi_2 |^2 + |\Psi_3|^2 ) = 1$. This rescaling redefines the self-interaction parameter and its corresponding corrections by multiplying them by the total number of particles $N$. For all simulations presented in this work, we considered $N = 4500$. This normalization of the total wave function allows us to derive a simple analytical expression for the chemical potential of the system in the stationary state (see the SM). 
\begin{figure}
    \centering
    \includegraphics[width=1.0\linewidth]{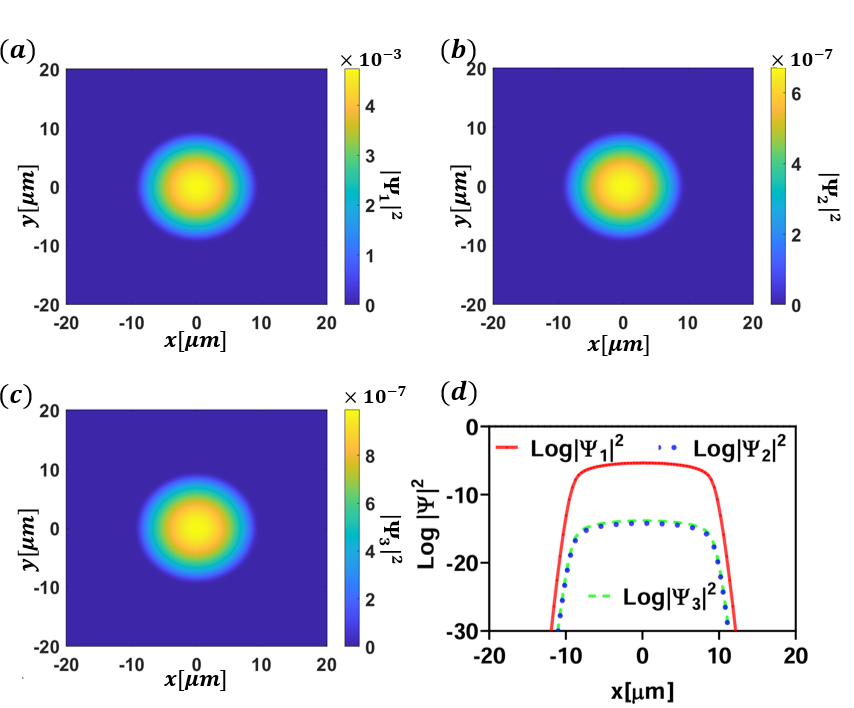}
    \caption{Stationary wave function for dipolar Bose-Einstein condensates. Population distribution in the $xy$ plane for the states (a) $\Psi_1$, (b) $\Psi_2$, and (c) $\Psi_3$. (d) Logarithm of the probability density of each condensate as a function of $x$.}
    \label{fig1}
\end{figure}

\begin{figure}
    \centering
    \includegraphics[width=1.05\linewidth]{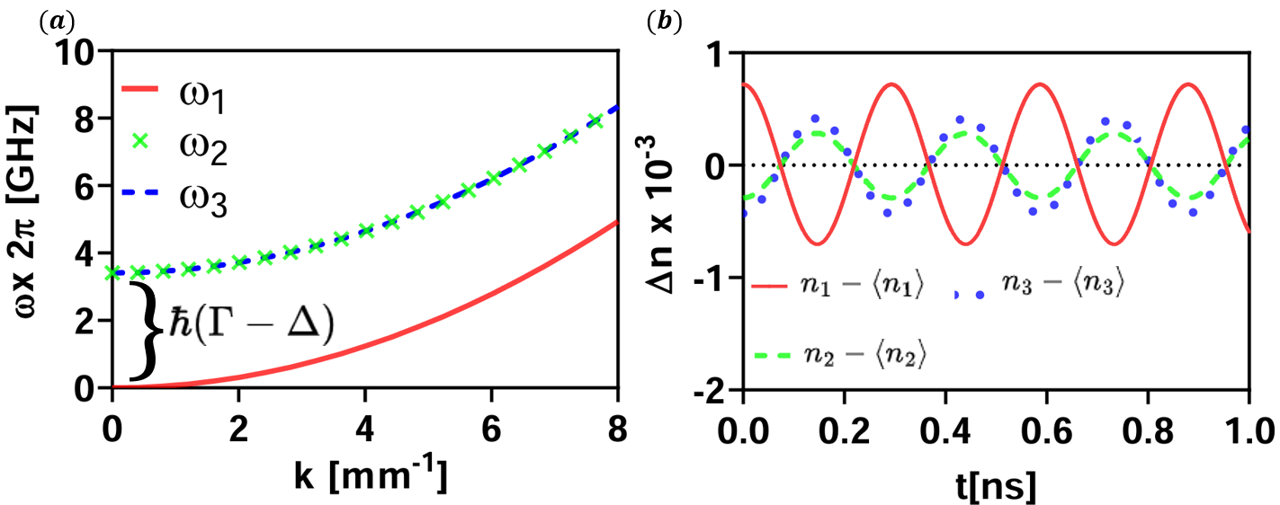}
    \caption{(a) Dispersion relation of the three condensates. (b) Population deviation from the mean as a function of time. $n_1$, $n_2$, and $n_3$ denote the populations of $\Psi_1$, $\Psi_2$, and $\Psi_3$, respectively, while $\langle n_1 \rangle $, $\langle n_2 \rangle$, and $\langle n_3 \rangle$ represent their corresponding mean values.}
    \label{fig2}
\end{figure}

\textit{Dipolar Bose-Einstein condensates}: to study the stationary and dynamical behavior of the system, we first consider the case where all three condensates are confined by the same harmonic trapping potential, i.e, $V = V_{t} \equiv \frac{1}{2}m \omega_{t}^2(x^2 + y^2)$. For this case, we observed Bose-Einstein condensation in all three components. This was achieved by solving  Eq.~(\ref{eq: Gross-Pitaevski}) using the imaginary time propagation (ITP) method \cite{Bao2004} to obtain the system's stationary configuration. The resulting relative populations were $n_1 = 0.9996$, $n_2 = 1.43 \times 10^{-4}$, and $n_3 = 2.13 \times 10^{-4}$, where $n_j = \int dr^2 |\Psi_j|^2$. As expected, the vast majority of particles occupy the ground state. In contrast, the excited states are only marginally populated, see figures \ref{fig1}.b and  \ref{fig1}.c.   The relative populations are primarily determined by the $\Gamma$ term and the pumping parameters. As shown in Fig. \ref{fig1}(d), the BEC wave function stabilizes into an approximately Gaussian profile, reflecting the influence of the harmonic trapping potential. Building on the stationary configuration in the absence of harmonic trap, we applied the Bogoliubov expansion \cite{Griffin2009,McClintock2011} around the steady-state solution, i.e., $\psi_j = e^{-i \mu t}\left( \psi_{jo} + u_j e^{-i \omega_j t} + v_j^* e^{i \omega_j t} \right)$, to derive the excitation spectrum, resulting in the dispersion relation shown at Fig. \ref{fig2}(a). We observe that the three condensates exhibit the typical dispersion relation of a BEC, with an energy shift of $\hbar (\Gamma-\Delta)$ in the upper condensates.

To study the dynamic behavior of the dipolar BEC, we solved the time-dependent Gross-Pitaevski equation using real-time evolution in the Crank-Nicolson scheme. As shown in Fig. \ref{fig2}(b), the dynamics is dominated by a Rabi-type process, in which the first condensate periodically feeds the other two condensates with a characteristic frequency of $3.46$ GHz. This value is consistent with the expected frequency $\sqrt{\Gamma ^2 +(\Omega_\pi^2 + \Omega_\sigma^2 )} \approx \Gamma$. It is noteworthy that populations of the condensates oscillate near their stationary values, and the upper condensates remain only marginally populated throughout the evolution.  
As observed in the stationary and dynamic behavior of the dipolar BEC, it comprises a primary condensate that holds nearly the entire population, along with two weakly populated secondary condensates coupled to it that induce small perturbations in the primary condensate.

\textit{Josephson effect in dipolar Bose-Einstein condensates}: to study the Josephson effect in the system, and taking into account the previous results, we performed an adiabatic elimination of the upper states, a well-established technique often used in the context of quantum optics \cite{Fewell2005,Brion2007}. As previously shown, the relative populations of the three condensates are extremely unbalanced, with the first condensate containing almost the entire particle population of the system. Even during the dynamics, the population ratios $\frac{|\psi_2(t)|^2}{|\psi_1(t)|^2}$ and $\frac{|\psi_3(t)|^2}{|\psi_1(t)|^2}$  remain of the order of $(\Omega_{\pi,\sigma}/\Gamma)^2$ at all times (see SM for the derivation), while the time derivatives $\dot{\psi}_2(t)$ and $\dot{\psi}_3(t)$ are of the order of $(\Omega_{\pi,\sigma}/\Gamma)^3$. This dynamical hierarchy justifies the adiabatic elimination of the excited components. With this in mind, we impose the following conditions on the upper states to ensure that their dynamics remain enslaved to the evolution of the ground state, i.e., $ 
    \dot{\Psi}_2 \approx 0 $ and $ \dot{\Psi}_3 \approx 0 $. This approximation is supported by the fact that the populations of the upper states oscillate rapidly around an equilibrium value. Therefore, on the timescales where we expect to observe the Josephson effect in the primary condensate, $\Psi_2$ and $\Psi_3$ can be considered quasi-stationary. Applying these approximations to the original GP equation, assuming the Thomas-Fermi condition for the upper states and considering that $\{{|\Psi_2|^2},{|\Psi_3|^2}\}\ll |\Psi_1|^2  $,  we obtain the following relations for $\Psi_2 $ and $\Psi_3$:
\begin{align}
    \Psi_2 =&\label{psi2ec} - \frac{\hbar \Omega_\pi e^{-i t \Delta }}{V+\hbar \Gamma + \gamma |\Psi_1|^2} \Psi_1\\
    \Psi_3 =&\label{psi3ec} - \frac{\hbar \Omega_\sigma e^{-i t \Delta }}{V+\hbar \Gamma + \gamma |\Psi_1|^2} \Psi_1.
\end{align}
Since $\hbar \Gamma$ is much larger than $V$ and $\gamma |\Psi_1|^2$, we can expand the expression and substitute it into the Gross-Pitaevskii equation for the primary condensate. Thereby, obtaining an effective GP equation for the wavefunction $\Psi_1$,
\begin{equation}
    i \hbar \partial_t \Psi_1 = \left[ - \frac{\hbar^2}{2m} \nabla^2 + V_{1{\rm eff}} + \gamma_{1{\rm eff}} |\Psi_1|^2\right] \Psi_1,
    \label{gppsi1}
\end{equation}
with the effective parameters given by,
\begin{align}
    V_{1{\rm eff}} &= V  + \frac{V}{\Gamma^2}(\Omega_\pi ^2  + \Omega_\sigma^2) - \frac{\hbar}{\Gamma}(\Omega_\pi ^2 + \Omega_\sigma ^2 )
    \label{veff},\\
    \gamma_{1{\rm eff}} &= \gamma\left( 1 + \frac{1}{\Gamma^2} (\Omega_\pi ^2 + \Omega_\sigma^2)\right).
    \label{gammaeff}
\end{align}
To validate this approximation, we derived the dispersion relations of the system using the effective model and compared them with those obtained from the original formulation, see SM. The resulting curves show excellent agreement, as demonstrated in Figure 2 of the SM. This approximation enables us to study the Josephson effect on the order of hertz at a frequency scale, which is fundamentally different from the Rabi dynamics that typically occur at gigahertz frequencies.

\begin{figure}
    \centering
    \includegraphics[width=1.0\linewidth]{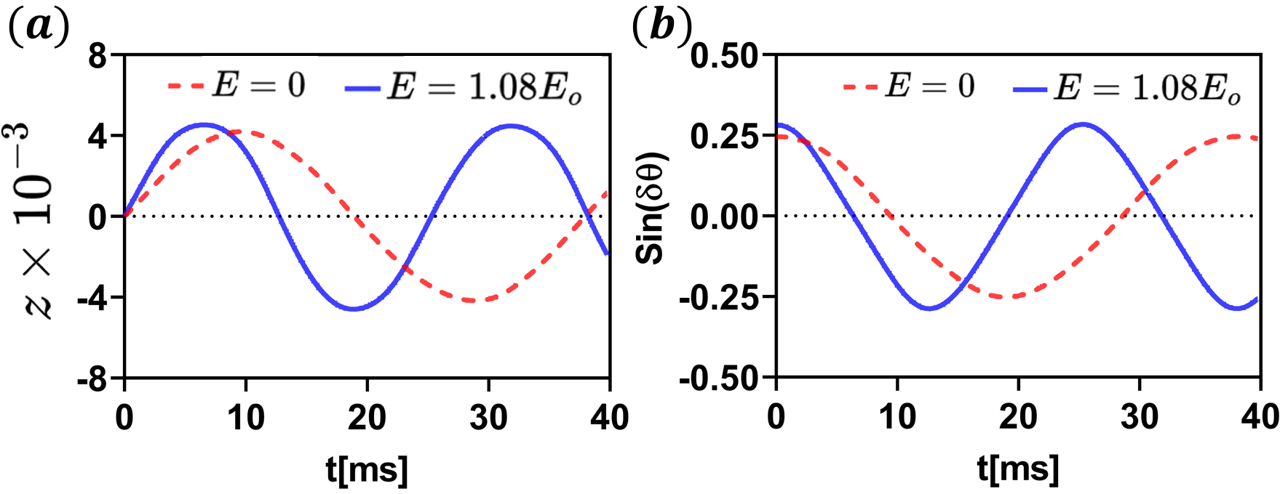}
    \caption{Josephson effect in dipolar condensates. (a) Fractional population imbalance $z$ and (b) sine of the relative phase $\delta \theta$ as functions of time under different electric field conditions. The red line corresponds to the case with no electric field, while the other curve displays the behavior under an electric field of $1.08 E_o$, with $E_o = 1.0\text{kV/cm}$.}
    \label{fig3}
\end{figure}

To induce Josephson dynamics in our system, we introduced a Gaussian potential barrier at the center of the harmonic trap that splits the condensate into two parts. The total potential is given by $V =V_t + V_{barrier} e^{- \frac{x^2}{2 \sigma^{2}}} $, where $V_{barrier} = 50 \hbar \omega_{t} $ and $\sigma = 0.4$ $\mu$m are the height and width of the Gaussian potential barrier, respectively. The chosen barrier height optimizes the tunability of the Josephson frequency by the applied electric field, whereas different barrier heights could reduce this range or even lead to self-trapping. It must be noted that the harmonic trap and the potential barrier act equally on the three condensates. We imposed a phase difference of ${\pi}/{2}$ between the two sides as an initial condition. The dynamical evolution are shown in Fig. \ref{fig3}, where we observed the characteristic Josephson oscillations in the fractional population imbalance and the relative phase, with a frequency of $26$ Hz.

Next, we apply an electric DC field $E$ in the $\hat{z}$ direction. Since we are working with polar NaCs molecules, the applied field acts as a Stark term in the quatum molecular Hamiltonian. Under this condition, the quantum states that describe the system are modified, giving rise to new dressed states. Therefore, our previous mapping from the states $\{| 0,1 \rangle, |1,0\rangle, | 1,1 \rangle  \}$ to the classical fields $\{ \Psi_1, \Psi_2, \Psi_3 \}$ is no longer the most appropriate framework for describing the system, as the effective electric dipole moment is not well defined in that basis. With this in mind, the first step is to obtain the new dressed states, which can be derived by diagonalizing the modified molecular Hamiltonian:
\begin{equation*}
    \hat{H}_{molecule} =\hat{H}_{Rot} +  \hat{H}_{DC}
\end{equation*}
Here, $\hat{H}_{Rot}$ denotes the rotational part of the molecular Hamiltonian, which in the basis of the $|J,M \rangle$ states is diagonal, with main elements of the form $BJ(J+1)$, where $B = \hbar \Gamma /2$ is the rotational constant of the NaCs molecule and $J$ is the angular momentum quantum number. The term $\hat{H}_{DC} = - E \hat{p}_z$ corresponds to the interaction with the external electric field, where $\hat{p}_z = p_o cos(\theta)$ is the dipole moment operator in spherical coordinates and $p_o$ is the permanent dipole moment\cite{Atkins2010-lg}. Further details are provided in the SM. 
\begin{figure}[h!]
    \centering
    \includegraphics[width=0.7\linewidth]{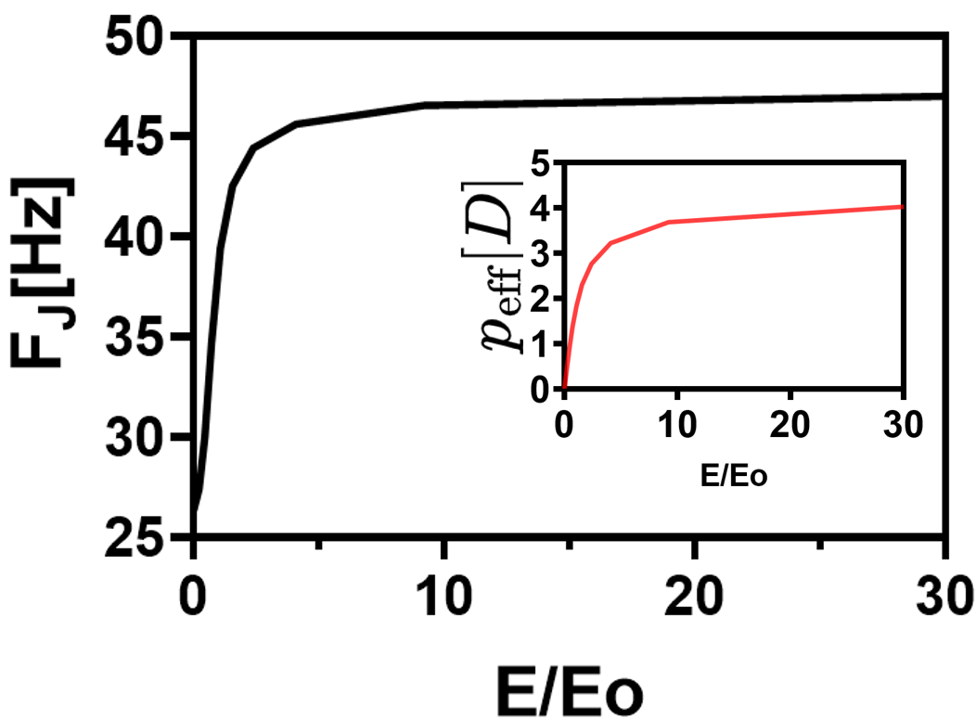}
    \caption{Josephson frequency as a function of the external electric field applied. $E_o = 1.0 \text{kV/cm}$. The inset shows the effective dipole moment as a function of the applied electric field.} 
    \label{fig4}
\end{figure}

From the diagonalization procedure, we obtain $n$ dressed eigenstates, where $n$ is the dimension of the Hilbert space. The applied electric field ensures that only one of these dressed states corresponds to the lowest-energy configuration. Since we perform a relaxation process prior to studying the Josephson dynamics, we define a new classical field, denoted as $\chi$, associated with this minimum-energy dressed state. Consequently, we introduce a new Gross-Pitaevskii equation governing the dynamics of this field:
\begin{equation*}
    i \hbar \dot{\chi} = \left[ - \frac{\hbar^2}{2m}\nabla^2 + V + \gamma |\chi|^2  + \Phi_{\chi} \right]\chi
\end{equation*}
For simplicity, we have assumed that the mass and the contact interaction term are not affected by the applied electric field, and that the Rabi and electric-field contributions are implicitly incorporated into the definition of the $\chi$ field. In the previous equation, $V$ denotes the external potential, which includes the harmonic trap and the central barrier, while $\Phi_{\chi}$ corresponds to the dipolar interaction term, given by:
\begin{equation*}
    \Phi_{\chi} =\int d\textbf{r}' \frac{\tilde{C}_{dd}}{4 \pi} \frac{1-3cos(\theta_{\chi})}{|\vec{r}-\vec{r}'|^3} |\chi (\vec{r}')|^2
\end{equation*}
It must be noted that $\tilde{C}_{dd} \propto p^2_{\text{eff}}$ depends on the effective dipole  moment $p_{\text{eff}}$ of the lowest-energy dressed state, and therefore on the $\chi$ field. Importantly, the value of this dipole moment varies with the applied electric field.

As shown in Fig. \ref{fig3} and Fig. \ref{fig4}, the Josephson frequency changes under the application of an external electric field, since both the effective dipole moment and, consequently, the dipolar interaction are modified. As the electric field increases, the effective dipole moment also increases, as illustrated in the inset of Fig. \ref{fig4} and reported in previous works\cite{Stevenson2023}, leading to an enhancement of the Josephson frequency. At sufficiently large electric fields, the effective dipole moment asymptotically approaches the permanent dipole moment of NaCs molecules, thereby stabilizing the Josephson frequency around $46 \text{Hz}$. In summary, the applied electric field tunes the Josephson frequency from 26 Hz at zero field to 46 Hz at high field, corresponding to an increase of approximately $77\%$. This pronounced tunability arises from the progressive enhancement of the effective dipole moment, which drives the system from a contact-dominated to a dipole-dominated Josephson regime. In real systems, Bose–Einstein condensates suffer one- and three-body losses that limit their lifetime. Our study considers the ideal stable case, but experiments have shown that microwave fields can suppress these losses, yielding lifetimes of 2 seconds \cite{Bigagli2024}, much longer than our millisecond simulation window, so the predicted dynamics should be observable. In laboratory conditions, Josephson oscillations may exhibit damping \cite{vanNieuwkerk2021,Lappe2018} or deflection, but the externally applied electric field remains experimentally controllable, allowing the oscillation frequency to be tuned within the damping-limited regime.

\begin{figure}
    \centering
    \includegraphics[width=0.7\linewidth]{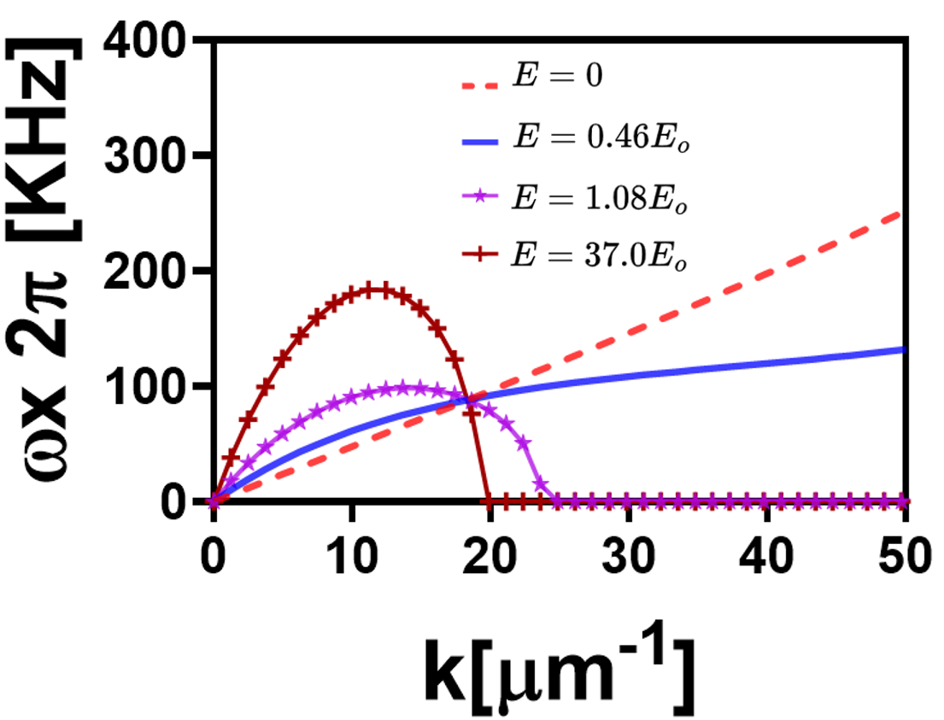}
    \caption{Dispersion relation of the system for the untrapped case at different electric field applied. The dashed line represent the system without electric field, the continous line represent $E=0.46 E_o$, the plus-line $E=1.08 E_o$ and the star-line $E = 37.0E_o$ with $E_o = 1.0 \text{kV/cm}$}
    \label{fig_disp_comparative}
\end{figure}

The Fig. \ref{fig_disp_comparative} displays the dispersion relations of the system under different applied electric fields. For strong electric fields, the system transitions into a fully dipolar regime, characterized by the appearance of typical roton minima \cite{Pendse2018} at intermediate field strenths and roton instabilities at high fields. The latter effect poses a challenge for stabilizing the condensate in the abscence of harmonic trap. Nevertheless, this highly dipolar behavior opens a pathway toward richer phases\cite{Bchler2007,Arazo2023,Mistakidis2024}, including pattern formation and even supersolidity under specific conditions.

It should be emphasized that, in this study, the applied electric field is oriented perpendicular to the particle distribution in the XY-plane. In this configuration, the system retains its planar symmetry, does not exhibit anisotropy, and the dipolar interaction term only contributes an additional repulsive pressure. By contrast, if an electric field with a component outside the z-direction were applied, the symmetry would be broken, inducing anisotropies. Such symmetry breaking could have major consequences for the Josephson oscillations.

\section{conclusions}

We studied the dynamical behavior of a Bose-Einstein condensate (BEC) exhibiting dipolar interactions. By aligning the dipole moments of the particles within our BECs and varying the parameters of the applied electric field, we systematically analyzed their impact on the system's nature and, consequently, on the Josephson junction properties, including the oscillation frequency. Specifically, our results reveal that by tuning the electric field, the system can transition between a typical Bose-Einstein condensate and a dipolar condensate, depending on the field strength. This transition significantly alters the system's properties, such as the  dispersion relation and sound velocity. Additionally, we show that the Josephson oscillation frequency can be finely tuned by adjusting the applied electric field, reflecting changes in dipole-dipole interaction energies. Our findings corroborate theoretical models that predict the sensitivity of the BEC to external fields, thereby enhancing our understanding of quantum coherence and tunneling in complex quantum systems.

\begin{acknowledgments}
Funding is acknowledged from DICYT regular 042431AP.  D. Gálvez-Poblete acknowledges ANID-Subdirección de Capital Humano/Doctorado Nacional/2023-21230818. A.S.N. and R.E.T. acknowledges funding from Fondecyt Regular 1230515 and 1230747, respectively .
\end{acknowledgments}

\end{document}